\documentclass[aps,prd,nofootinbib,showpacs,preprintnumbers,longbibliography,amssymb,floatfix,11pt]{revtex4-1} 

\usepackage[sort&compress]{natbib}
\usepackage[dvipsnames]{xcolor}

\usepackage{graphicx}
\usepackage{epsfig}
\usepackage{dcolumn}
\usepackage{bm}

\usepackage{amsmath}
\usepackage{siunitx}
\usepackage{amssymb}
\usepackage{physics}
\usepackage{slashed}
\usepackage{dsfont}
\usepackage{subcaption}
\usepackage{hyperref}
\usepackage{cleveref}
\usepackage{soul}

\hypersetup{
    colorlinks=true,       
    linkcolor=purple,   
    citecolor=blue,        
    filecolor=magenta,      
    urlcolor=blue           
}

\graphicspath{{Figures/}}

\begin{document}

\preprint{MITP-23-057}

\title{Consistent Electroweak Phenomenology of a Nearly Degenerate $Z'$ Boson}

\author{Prisco Lo Chiatto}
\email{plochiat@uni-mainz.de}
\author{Felix Yu}
\email{yu001@uni-mainz.de}
\affiliation{PRISMA+ Cluster of Excellence \& Mainz Institute for Theoretical Physics, Johannes Gutenberg University, 55099 Mainz, Germany}

\begin{abstract}
Extracting constraints on the kinetic mixing of a new $U(1)'$ gauge boson hiding under the Standard Model $Z$ boson resonance requires the formalism of non-Hermitian two-point correlation functions at 1-loop order. We derive self-consistent collider constraints on $Z'$ bosons with kinetic mixing in a narrow mass window around the $Z$ boson, considering both model-independent and model-dependent bounds. Our treatment elucidates the importance of both avoided level crossing and the dispersive seesaw effect in interpreting the existing constraints. We also discuss the implications for future measurements on the $Z$-pole.
\end{abstract}

\maketitle


\section{Introduction}
\label{sec:Introduction}

New $Z'$ gauge bosons are a commonly studied feature of many beyond the Standard Model (BSM) theories, including new $U(1)'$ gauge symmetry extensions of the Standard Model (SM)~\cite{Carena:2004xs,  Batra:2005rh, Langacker:2008yv, Hook:2010tw, FileviezPerez:2011pt, Dobrescu:2013cmh, Dobrescu:2014fca, Costa:2019zzy, Heeba:2019jho, Michaels:2020fzj, Bauer:2020itv}, models with a vector or axial-vector portal to dark matter or dark sectors~\cite{LHCNewPhysicsWorkingGroup:2011mji, Liu:2017lpo, Dror:2017nsg, Albert:2017onk, Ismail:2017ulg, Ismail:2017fgq, Dror:2018wfl, Foldenauer:2018zrz, Cheng:2024hvq}, and models with light vector dark matter~\cite{Chu:2011be, Essig:2013lka, Ilten:2018crw, Bauer:2018onh, Fabbrichesi:2020wbt, Antel:2023hkf}.  Typically, these new $Z'$ vectors are assumed to be well-separated in mass from existing neutral vector states in the SM, including both the $Z$ boson of electroweak symmetry as well as vector mesons from quantum chromodynamics (QCD).  Correspondingly, for weakly-coupled $Z'$ bosons, the narrow-width approximation affords an intuitive factorization of the BSM signal phenomenology from the SM background.

Importantly, the narrow-width approximation and subsequent factorization can be improved by increasing theoretical precision, as in the well-studied case of the $Z$-$\gamma$ mixing calculation necessary to extract $Z$-pole observables at the Large Electron-Positron (LEP) experiments~\cite{Jegerlehner:1990uiq, ALEPH:2005ab}.  Specifically, the large mass difference between the $Z$ and the photon affords a sound Dyson resummation of the 1-particle irreducible two-point correlation function, given the inverse of the fractional mass separation is much smaller than $\alpha$, so that interference effects can be treated as subleading at each order in perturbation theory.  

In the context of a $U(1)'$ gauge symmetry extension to the SM, the $Z'$ mass parameter and a kinetic mixing coupling~\cite{Holdom:1985ag} between the $Z'$ field strength and the SM hypercharge field strength control the vector boson mass spectrum.  Given that the $Z$ width is measured to be $2.4952 \pm 0.0023$~GeV~\cite{ParticleDataGroup:2024cfk}, there is an approximate $5$~GeV window for input $Z'$ masses where a cross section factorization according to the Breit-Wigner prescription~\cite{Breit:1936zzb} breaks down.  In this mass range, large kinetic mixing between the $Z'$ and $Z$ boson is expected to be completely excluded given that the $Z'$ properties are expected to wildly disrupt $Z$-pole observables.  This intuition is imprecise, however, since there must exist a decoupled regime where kinetic mixing becomes negligible and $Z'$ bosons avoid exclusion from LEP data.

Here, we employ the required framework to address this oversight, which enables us to calculate the collider phenomenology of $Z'$ bosons in a mass range close to the $Z$ boson and to rigorously extract limits on the kinetic mixing and other $Z'$ properties.  Clearly, this framework is highly relevant for interpreting correctly the existing LEP data.  Our work is also motivated by the possibility of future $Z$-pole factories, such as FCC-ee~\cite{FCC:2018evy} and CEPC~\cite{CEPCStudyGroup:2018ghi}, which will test the SM electroweak parameters to unprecedented precision and can offer singular sensitivity in the kinetic mixing vs.~$Z'$ mass plane.  Moreover, studies of other $Z'$ masses degenerate with SM vector mesons are then also tractable with our framework.

Our calculation is based on the work developed in Refs.~\cite{Frank:2006yh, Fuchs:2016swt}. This treatment of 1-loop mixing and oscillation has been adapted to constrain extra Higgs scalars~\cite{Fuchs:2016swt} as well as right-handed neutrinos~\cite{Racker:2020avp, Racker:2021kme}. In hindsight, this formalism also addresses standard studies of $K_L$, $K_S$ mixing and oscillation as well as neutrino mixing and oscillation~\cite{Beuthe:2001rc}.  Earlier work on particle mixing in the context of resonant CP violation for strongly mixed scalar or fermion states includes Ref.~\cite{Pilaftsis:1997dr}.  The collider phenomenology of two BSM vector bosons with nearly degenerate masses has been considered in Refs.~\cite{Cacciapaglia:2009ic, deBlas:2012qp}; in this paper, we instead consider adding one single vector boson to the SM, and letting it be nearly degenerate with the SM $Z$ resonance.

We will specifically focus on $Z$, $Z'$, and $\gamma$ vector boson mixing with the $Z$ and $Z'$ bosons being closely spaced in mass.  Since we are motivated to understand very weakly-coupled $Z'$ bosons, we will remain agnostic about the explicit current mediated by the $Z'$ boson and treat the kinetic mixing parameter independently of the $Z'$ gauge coupling.  In practice, this is afforded by a hierarchical structure of heavy fermions coupled to the $Z'$ boson that satisfy a trace condition which renders the kinetic mixing parameter to be logarithmic and suppressed by a loop factor~\cite{Dobrescu:2021vak}.  We also recognize that interacting massive $Z'$ bosons are generally restricted to a limited energy range of validity from unitarity unless the $U(1)'$ symmetry is Higgsed~\cite{Kribs:2022gri}, but the $U(1)'$ Higgs sector plays no role in our discussion and will also be neglected.

In comparison to earlier treatments of both mixing and oscillation in quantum field theory observables as well as earlier treatments of kinetic mixing for vector bosons, this work will emphasize the mathematical necessity for non-unitary transformations needed to diagonalize the kinetic mixing as well as the independent diagonalization of the two-point correlation functions and their corresponding impact on phenomenology. This will illustrate fundamental properties of $U(1)$ gauge theories in relation to the optical theorem and the required non-Hermiticity of unstable state masses.

Previous work studying the effects of $Z'$ bosons on electroweak observables include Refs.~\cite{Babu:1997st, Hook:2010tw, Curtin:2014cca, Greljo:2022dwn, Harigaya:2023uhg, Qiu:2023zfr}.  Notably, only Refs.~\cite{Hook:2010tw} and~\cite{Qiu:2023zfr} explicitly consider the complications of strong $Z$-$Z'$ mixing in the quasi-degenerate regime, and a related recent discussion of large $Z'$ mixing with the $\rho$ meson was presented in~\cite{Coyle:2023nmi}.

Our main goal in this work is to present consistent constraints on kinetic mixing in the quasi-degenerate region around the $Z$ pole, calculated from three primary observables: the $W$ boson mass, the $Z$ boson width and the muon pair production rate.  We emphasize the importance of avoided level crossing in interpreting the available parameter space, as well as the correct assignment of 1-loop widths that scale nontrivially with regards to the kinetic mixing parameter and $Z$-$Z'$ mass splitting, which reflects the decoupling of the $Z'$ boson in the double limit of both vanishing kinetic mixing and mass splitting.  Unfortunately, previous work did not correctly calculate the physics of the double limit and hence consistent limits on kinetic mixing in the quasi-degenerate regime are presented here for the first time.  We critique the earlier literature in~\cref{sec:doublelimit}.  The novelty of our results stemming from the phenomena of avoided crossing and the dispersive seesaw effect offers a fresh view of the physics of global symmetries, since these phenomena are a characteristic feature of the quantum physics of two-level systems.

In~\cref{sec:Framework}, we present the formalism for deriving the collider phenomenology of the $Z$, $\gamma$ and a new $Z'$ boson.  We derive the improved Breit-Wigner propagators, which reflect the correct pole structure of the massive vector states at 1-loop order.  In~\cref{sec:Model}, we discuss the essential features of our $Z'$ model and the canonical diagonalization of kinetic mixing.  We then expand on the quasi-degenerate regime, explaining highly-mixed but decoupled properties of the $Z'$ boson via the dispersive seesaw effect and review the feature of avoided crossing.  In~\cref{sec:results}, we present the collider constraints of a quasi-degenerate $Z'$ boson in the kinetic mixing vs.~mass plane.  We conclude in \cref{sec:conclusions}.  In~\cref{sec:doublelimit}, we present the physics of the double limit and critically evaluate the previous literature that analyzed the quasi-degenerate regime.


\section{Modified Breit-Wigner Propagators for the Quasi-Degenerate Case}
\label{sec:Framework}

The 1-loop mixing between the SM $Z$ and $\gamma$ is a canonical example where extraction of constraints on new physics from LEP $Z$-pole observables using the simple tree-level factorization by the $Z$ mass and its branching fractions~\cite{Bohm:1986rj} does not lead to the desired level of precision. Indeed, especially when considering new physics affecting electroweak parameters~\cite{Peskin:1990zt, Peskin:1991sw}, $\gamma-Z$ mixing, which arises at 1-loop, has to be included. When particles have flavor interactions that are misaligned with the mass basis, as for the case of the active neutrinos in the SM, the phenomenon of oscillation is also observed.  Mixing and oscillation phenomena simultaneously occur in neutral kaon propagation, where the 1-loop mixing in the mass eigenstates $K_L$ and $K_S$ and their oscillating decay amplitudes gave the first direct evidence for $CP$ violation in weak interactions~\cite{Christenson:1964fg}.  In the $Z$, $Z'$, and $\gamma$ case, however, since the $Z$ boson's lifetime is $\sim 10^{-25}$ s, oscillation phenomena are completely unobservable at present and future colliders.  Hence, we will focus on the modification of the $Z$ propagator and decay rates.  The question of whether oscillation might be phenomenologically relevant for realistic BSM vector bosons nearly degenerate with QCD vector resonances with larger lifetimes, such as the $\Upsilon$, is left for future work.


Following the analyses of Refs.~\cite{Frank:2006yh, Fuchs:2016swt}, we provide a modified Breit-Wigner approach to the 1-loop mixing of quasi-degenerate vector bosons.  We begin by remarking that since we will be focused on contributions of massive vector bosons near their poles, we can safely neglect the longitudinal parts of the vector boson propagators.  This is justified since near the pole, only the pole mass and width, both gauge-invariant quantities, contribute non-negligibly to observables~\cite{Cacciapaglia:2009ic}.  We remark this consideration also side-steps questions of gauge invariance and unitarity in the renormalization of gauge bosons~\cite{Stuart:1991cc, Sirlin:1991rt, Sirlin:1991fd, Nowakowski:1993iu, Denner:2014zga}.\footnote{In particular, the full propagator of a vector boson, in $R_\xi$ gauge, is expressed as 
\begin{equation}
i \Delta_{\mu \nu} = 
\left( g_{\mu\nu}-\frac{p_\mu p_\nu}{p^2} \right) \frac{-i}{p^2-m^2+\Pi_T} + \frac{p_\mu p_\nu}{p^2} \frac{-i\xi}{p^2-\xi(m^2-\Pi_T-p^2\Pi_L)} \ , 
\nonumber
\end{equation}
where $\Pi_T$ and $\Pi_L$ are defined via 
$\Pi_{\mu \nu} = \Pi_T g_{\mu \nu} + \Pi_L \, p_\mu p_\nu$,
with $\Pi_{\mu \nu}$ denoting the self-energy at momentum $p$ ~\cite{Cacciapaglia:2009ic}.
We remark that both the gauge-dependent term -- which will anyway cancel against the ghost contributions in observables ~\cite{Jegerlehner:1990uiq} -- and the gauge-independent term contributing to the longitudinal propagator are numerically negligible, since they are suppressed by the small ratio $m_f/\sqrt{p^2}$, with $m_f$ (typically the electron mass) as the smaller of the initial and final fermion masses, after contracting the propagator with the incoming and outgoing fermions.}

Compared to the standard Breit-Wigner (BW) approximation, the main complication in our scenario is that we study mass splittings that are parametrically comparable or smaller than the leading order widths of the involved particles.  As a result, it is not possible to assign unequivocally a given field to a pole of the 2-pt.~Green's function.  Moreover, the standard approach of resumming 1-particle irreducible diagrams into a single propagator presumes large mass splitting from other asymptotic states, and trying to recover a $\delta m \to 0$ limit leads to off-diagonal mixing contributions that are beyond perturbative control, such that a modification of the BW prescription is needed: an early construction is found in Ref.~\cite{Feshbach:1958nx}. 

Separately, propagators of unstable particles treated in an on-shell formalism should mix orders of perturbation theory, as required by the optical theorem.  In a BW approximation, only a subset of diagrams are retained in the Dyson series resummation, which generally sacrifices gauge invariance and also unitarity of the theory at high energies~\cite{Stuart:1991cc, Sirlin:1991rt, Sirlin:1991fd, Denner:2014zga}.  One of the advantages of the 1-loop treatment of Ref.~\cite{Fuchs:2016swt} is the inclusion of interference diagrams in the Dyson resummation, which improves the treatment of gauge invariance and hence ameliorates unitarity issues. As a result, since only gauge invariant quantities are used in the calculation, we retain good accuracy in a wide regime around the pole masses.

We begin with the renormalized irreducible 2-pt.~vertex function matrix as a function of the Mandelstam variable $s$.  Denoting the vector fields as $Z$, $A$, and $K$ for the SM $Z$ boson, photon, and new physics gauge boson, respectively, we have
\begin{equation}
  \mathbf{\Gamma} = i (s\mathds{1} - \mathbf{M}(s)) = i
\begin{pmatrix}
  s - m_{Z}^2 - \Pi^T_{ZZ}(s) & - \Pi^T_{ZA}(s) & - \Pi^T_{ZK}(s)\\
  -\Pi^T_{ZA}(s) & s - \Pi^T_{AA}(s) & - \Pi^T_{AK}(s)\\
  -\Pi^T_{ZK}(s) & -\Pi^T_{AK}(s) & s - m_{K}^2 - \Pi^T_{KK}(s)
  \end{pmatrix}
  \label{eqn:GammaMat}
\end{equation}
where $\mathbf{M}(s)$ is the 1-loop improved mass squared matrix that contains the renormalized self-energies.  The propagator matrix is the negative inverse of $\mathbf{\Gamma}$: $\mathbf{\Delta}(s) = - \left( \mathbf{\Gamma} \right)^{-1}$.

As mentioned previously, we only retain the transverse part of the propagator as we are only interested at the region near the pole where the (gauge-dependent) longitudinal part of the propagator is numerically negligible.  Furthermore, from now on, we will not write down the $s$ dependence or the $T$ superscript for the self energies $\Pi_{ij}$, keeping them implicit. 

We first consider the diagonal entries of the propagator matrix $\mathbf{\Delta}$, which are constructed from $\mathbf{\Gamma}$ entries as~\footnote{In this and following expressions, repeated indices are not summed over.}
\begin{equation}
\Delta_{ii}(s) = \frac{\Gamma_{jj}\Gamma_{kk}-\Gamma_{jk}^2}{\Gamma_{ii}\Gamma_{jk}^2 + \Gamma_{jj}\Gamma_{ki}^2+\Gamma_{kk}\Gamma_{ij}^2 -\Gamma_{ii}\Gamma_{jj}\Gamma_{kk} - 2\Gamma_{ij}\Gamma_{jk} \Gamma_{ki}} \equiv \frac{i}{s-m_{i}^2+\Pi_{i}^{\text{eff}}} \ ,
\label{eqn:Deltaii}
\end{equation}
where $\Gamma_{ij}$ is the $(ij)$ entry of $\mathbf{\Gamma}$, $i = 1$, $2$, $3$; $j, k \neq i$, and we introduced the effective self-energy 
\begin{equation}
  \Pi_{i}^{\text{eff}} = \Pi_{i} - i \frac{\Gamma_{ij}\Gamma_{jk}\Gamma_{ki}-\Gamma_{ki}^2\Gamma_{jj}-\Gamma_{ij}^2\Gamma_{kk}}{\Gamma_{jj}\Gamma_{kk}+\Gamma_{jk}^2} \ ,
\label{eqn:Pieffii}
\end{equation}
which preserves the form of an unmixed propagator while including $3 \times 3$ mixing contributions, with $\Pi_1 = \Pi_{ZZ}$, $\Pi_2 = \Pi_{AA}$, and $\Pi_3 = \Pi_{KK}$. Note that only the first term of \cref{eqn:Pieffii} survives in the zero mixing limit.

Next, the off-diagonal entries of $\mathbf{\Delta}$ are
\begin{equation}
  \Delta_{ij}(s) = \frac{\Gamma_{ij}\Gamma_{kk} - \Gamma_{jk}\Gamma_{ki}}{\Gamma_{ii}\Gamma_{jj}\Gamma_{kk}+2\Gamma_{ij}\Gamma_{jk}\Gamma_{ki}-\Gamma_{ii}\Gamma_{jk}^2-\Gamma_{jj}\Gamma_{ki}^2-\Gamma_{kk}\Gamma_{ij}^2} \ .
\label{eqn:Deltaij}
\end{equation}

The poles of the characteristic polynomial of $\mathbf{\Delta}$ are by definition zeros of $\det(\mathbf{\Gamma})$. This observation allows us to rewrite the elements of $\mathbf{\Delta}$ as 
\begin{equation}
  \Delta_{ii} = \frac{(s-m^2_{j}+\Pi_{jj})(s-m^2_{k}+\Pi_{kk}) - \Pi_{jk}^2}{\det(\mathbf{\Gamma})} \ ,
\quad 
  \Delta_{ij} = \frac{\Pi_{jk}\Pi_{ki} - \Pi_{ij}(s-m^2_{k}+\Pi_{kk})}{\det(\mathbf{\Gamma})} \ .
  \label{eqn:FullPropagators}
  \end{equation}
This form makes it clear that \textit{each} of the elements of the propagator matrix has as many poles as zeros of $\det(\mathbf{\Gamma})$, \textit{i.e.}~three. This suggests to approximate the propagators as the sum of three Breit-Wigner propagators,
\begin{equation}
  \Delta_{ij} \approx \sum_{a=I}^{III} \mathbf{Z}^{ai}\mathbf{Z}^{aj} \frac{i}{s-\mathcal{M}^2_a} \equiv \sum_{a=I}^{III} \mathbf{Z}^{ai}\mathbf{Z}^{aj}\Delta^{\text{BW}}_a  \ , \text{ for } i, j = 1, 2, 3 \ ,
\label{eqn:ModifiedBWPropagator}
\end{equation}
in order to maintain the correct pole structure, where $\mathbf{Z}_{ai},\mathbf{Z}_{aj}$ are constant coefficients to be determined and the poles are at $\mathcal{M}_a^2$, which can be found by solving $\det (\mathbf{\Gamma}) = 0$.

The coefficients in \cref{eqn:ModifiedBWPropagator} are determined by imposing the renormalization conditions of unit residue and vanishing mixing on-shell, {\it i.e.}~we construct a matrix $\mathbf{Z}$ such that
\begin{align}
\lim\limits_{s \to \mathcal{M}_{I}^2} \dfrac{-i}{s-\mathcal{M}_I^2} \left( \mathbf{Z}\Gamma \mathbf{Z}^T \right)_{11} &= 1 \ , \\
\lim\limits_{s \to \mathcal{M}_{II}^2} \dfrac{-i}{s-\mathcal{M}_{II}^2} \left( \mathbf{Z}\Gamma \mathbf{Z}^T \right)_{22} &= 1 \ , \nonumber \\
\lim\limits_{s \to \mathcal{M}_{III}^2} \dfrac{-i}{s-\mathcal{M}_{III}^2} \left( \mathbf{Z}\Gamma \mathbf{Z}^T \right)_{33} &= 1 \ , \nonumber
\label{eqn:RenConds}
\end{align}
and
\begin{equation}
\lim\limits_{s \to \mathcal{M}_a^2} \dfrac{-i}{s-\mathcal{M}_a^2}\left( \mathbf{Z}\Gamma \mathbf{Z}^T \right)_{ij} = 0 \ ,
\end{equation}
for every $a =I,II,III, i\neq j$. 

For example, by expanding the diagonal entry propagator $\Delta_{11}$ around the pole $s = \mathcal{M}^2_I$, we obtain 
\begin{equation}
\Delta_{11}\approx \frac{i}{s-\mathcal{M}^2_I} \frac{1}{1+\partial_s\Pi_{i}^{\text{eff}}(s=\mathcal{M}^2_I)} \equiv \Delta_I^{\text{BW}} Z_{I1}^2 \ ,
\label{eqn:Exp11}
\end{equation}
where the entry 
\begin{equation}
    Z_{I1} = \frac{1}{ \sqrt{ 1 + \partial_s \Pi_1^{\text{eff}} (s = \mathcal{M}_I^2)}} \ .
        \label{eqn:ZDiag}
\end{equation}
On the other hand, for the off-diagonal entry $\Delta_{12}$, expanding around $s = \mathcal{M}^2_I$ gives 
\begin{equation}
    \Delta_{12} = \Delta_{11} \dfrac{\Delta_{12}}{\Delta_{11}} \approx \left . (Z_{I1})^2 \Delta_I^{\text{BW}}  \dfrac{\Delta_{12}}{\Delta_{11}}  \right|_{s = \mathcal{M}^2_I} \mkern-36mu \equiv (Z_{I1})^2R_{I2}\Delta_{I}^{\text{BW}}\, .
    \label{eqn:ZOffDiag}
\end{equation}
The $\mathbf{Z}$ matrix then is
\begin{equation}
\mathbf{Z} =
\begin{pmatrix}
Z_{I1} & Z_{I1} R_{I2} & Z_{I1} R_{I3} \\
Z_{II\,2} R_{II\,1}& Z_{II\,2} & Z_{II\,2} R_{II\,3} \\
 Z_{III\,3} R_{III\,1} & Z_{III\,3} R_{III\,2} & Z_{III\,3} \\
\end{pmatrix}\, ,
\end{equation}
with elements defined in analogy with~\cref{eqn:ZDiag,eqn:ZOffDiag}.
Note that the entries of the $\mathbf{Z}$ matrix can also be interpreted as the finite wave-function renormalization factors required by the Lehmann-Symanzik-Zimmermann (LSZ) formalism to be able to employ propagators for external particles~\cite{Fuchs:2016swt, Espriu:2002xv}.\footnote{The LSZ formalism for mixed states has been analyzed in detail in Refs.~\cite{Lewandowski:2017omt, Lewandowski:2018bnn}.}

We can now write the momentum-space amplitude for a process $X \to Y$ mediated by $s$-channel exchange of mixing gauge bosons~\cite{Fuchs:2016swt}
\begin{align}
  \mathcal{A} (X \to Y) &= \sum_{i,j = 1,2,3} V_{i}^{X} \Delta_{ij}(s) V_{j}^{Y} \approx  \sum_{i,j = 1,2,3} V_{i}^{X} \left(\sum_{a = I}^{III} \mathbf{Z}_{ai}\Delta_{a}^{\text{BW}}(s) \mathbf{Z}_{aj}\right)V_{j}^{Y} 
  \nonumber \\
	      & =  \sum_{a=I}^{III} V_{\text{eff},\,a}^X \Delta_{a}^{\text{BW}}(s) V_{\text{eff}, a}^{Y} \ , 
\label{eqn:AmplitudeBW}
\end{align}
where we defined the effective vertex functions $V_{\text{eff}}$ as the sum of the vertex functions of the flavor eigenstates weighted by the $Z$ factors, that is
\begin{equation}
V_{\text{eff},\,a}^X = \sum_{i=1,2,3} \mathbf{Z}_{ai}V^X_{i} \ ,
\label{eqn:GammaEff}
\end{equation}
and $V_i^X$ is the vertex function of the given in-state in momentum space, {\it i.e.}~the Feynman rule.

If the BSM couplings are kept small, they mostly affect the vertex functions, as expected. However, the off-diagonal propagators in~\cref{eqn:ModifiedBWPropagator} can be comparable with the diagonal propagators for any BSM coupling, if the mass splitting is small enough.

Armed with~\cref{eqn:GammaEff}, in~\cref{sec:results} we will study the collider phenomenology of the general kinetic mixing that we introduce in the next section. We should stress that we only consider the 1-loop contributions to the propagator of the vector bosons, and ignore the 1-loop vertex renormalization, as we are interested in the phenomenology near the $Z$ pole, where non-resonant contributions are heavily suppressed.

\section{Quasi-Degenerate Vector Bosons with Kinetic Mixing, the Dispersive Seesaw Effect and Avoided Crossing}
\label{sec:Model}

In this section, we review the procedure to analyze the spectrum and couplings of a $U(1)'$ gauge symmetry that is kinetically mixed with SM hypercharge, initially following Ref.~\cite{Liu:2017lpo}.  As mentioned in the Introduction, we will not need to consider Higgs mixing effects in regard to the treatment at the poles of the degenerate $Z$ and $Z'$ bosons.  We will also treat the $Z'$ current as generic but leave all SM matter fields uncharged under $U(1)'$.  After outlining the standard solution for solving the kinetic mixing, we analyze the case when the $Z$ and $Z'$ bosons are quasi-degenerate.  The phenomenological consequence of the propagator treatment of the $Z$ and $Z'$ boson presented in~\cref{sec:Framework} results in a dispersive seesaw effect, which is introduced in~\cref{subsec:ToyModel} and applied in~\cref{subsec:QZapplied}, that drives the widths apart and, counter to the naive expectation that the $Z$ and $Z'$ bosons are highly mixed when quasi-degenerate, implies decoupling of the $Z'$ from the SM.  The phenomenon of avoided crossing is also reviewed.  In~\cref{sec:doublelimit}, we critically evaluate the previous literature discussing the quasi-degenerate $Z$-$Z'$ regime and point out a flaw in previous work that led to incorrect calculations of the bounds in the kinetic mixing vs.~$Z'$ mass plane.

We denote the {\it gauge basis} field $K$ for the new $U(1)'$ symmetry gauge boson when the Lagrangian includes the kinetic mixing between the hypercharge and $K$ field strength tensors, reserving the $Z'$ notation when discussing the {\it mass basis} new physics vector field.

We start with the relevant terms of the Lagrangian after spontaneous symmetry breaking $SU(2)_L \times U(1)_Y \times U(1)' \to U(1)_{\text{EM}}$,
\begin{equation}
  \mathcal{L} \supset -\frac{1}{4}B_{\mu\nu}B^{\mu\nu} - \frac{1}{4}W_{\mu\nu}^iW^{i\,\mu\nu}-\frac{1}{4}K_{\mu\nu}K^{\mu\nu} + \frac{\chi}{2 \cos(\theta_W)}B_{\mu\nu}K^{\mu\nu} + \frac{1}{2} 
  \begin{pmatrix}
    W^{3\,\mu} & B^\mu& K^\mu
  \end{pmatrix}
  \mathbf{M}
  { \begin{pmatrix}
      W_{\mu}^3 \\ B_\mu\\ K_\mu
  \end{pmatrix}} \ ,
  \label{eqn:LagKinMix}
\end{equation}
where $\mathbf{M}$ is the tree-level mass squared matrix of the vector fields in the gauge basis, $\chi$ is the kinetic mixing parameter, and $\theta_W = \arctan (g' / g)$ is the SM weak mixing angle.  As detailed in Ref.~\cite{Liu:2017lpo}, we perform a weak angle rotation $R_W$ which diagonalizes the mass matrix but apportions the hypercharge kinetic mixing between the SM $Z$ and $\gamma$ states.  The diagonalization of the kinetic mixing is accomplished by subsequent transformations $U_1$ and $U_2$, which diagonalize the kinetic mixing but reintroduce mass mixing between the $Z$ and $K$ vector states.  A final mass matrix rotation $R_M$ is performed to diagonalize the vector masses.  The corresponding matrices are
\begin{align}
R_W &= 
  \begin{pmatrix}
    c_W & s_W & 0\\
    -s_W & c_W & 0\\
    0 & 0 & 1
  \end{pmatrix}\,, \ 
    U_{1} = 
\begin{pmatrix}
  1 & 0 & 0 \\
  -\chi^2 t_{W} & 1 & \chi \\
  -\chi t_{W} & 0 & 1
  \end{pmatrix}\,, \ 
  U_{2}=
\begin{pmatrix}
  \sqrt{\frac{1-\chi^2}{1-\chi^2c_W^{-2}}} & 0 & 0 \\
  0 & 1 & 0 \\
  \frac{-\chi^3 t_{w}}{\sqrt{(1-\chi^2)(1-\chi^2c_{W}^{-2})}} & 0 & \frac{1}{\sqrt{1-\chi^2}}
  \end{pmatrix} \ , 
  \label{eqn:allRots}
  \\
R_M &= 
  \begin{pmatrix}
    c_M & 0 & s_M \\
    0 & 1 & 0 \\
    -s_M & 0 & c_M 
  \end{pmatrix} \,, \
\tan \theta_M = \dfrac{1}{\beta \pm \sqrt{\beta^2 + 1}} \,,\
\beta \equiv \dfrac{m_{Z,\text{ SM}}^2(1 - \chi^2)^2 - m_K^2 (1 -\chi^2 \, c_W^{-2} - \chi^2 \, t_W^2)}{2m_K^2 \, \chi \, t_W\sqrt{1-\chi^2 \, c_W^{-2}}} \ ,
\nonumber
\end{align}
where the upper (lower) sign in the definition of $\theta_M$ corresponds to $m_{Z,\text{ SM}} > m_K \, (m_{Z,\text{ SM}} < m_K)$.

After diagonalization, this tree-level vector boson mass squared matrix reads 
\begin{equation} 
\text{diag}
\left( \frac{m_{Z,\text{ SM}}^2 + m_K^2 - \chi^2 \, m_{Z,\text{ SM}}^2 + \Delta_\chi} {2(1-\chi^2 \, c_W^{-2})},~ 0, ~ \frac{m_{Z,\text{ SM}}^2 + m_K^2-\chi^2 \, m_{Z,\text{ SM}}^2 - \Delta_\chi} {2(1 - \chi^2 \, c_W^{-2})}\right) \ ,
\end{equation}
with
\begin{equation}
\Delta_\chi =
\sqrt{(m_{Z,\text{ SM}}^2\left( 1-\chi^2 \right)-m_K^2)^2 + 4 m_{Z, \text{ SM}}^2 m_{K}^2 \, \chi^2 \, t_W^2} \ .
\label{eqn:MassEig}
\end{equation}
We remark that the massive gauge bosons exhibit avoided crossing already at tree-level, since $\Delta_\chi$ always causes the mass eigenvalues to move further apart compared to the diagonal entries before the $\theta_M$ rotation.  In summary, the new mass basis for the neutral gauge bosons is
\begin{equation}
\begin{pmatrix}
  \tilde{Z}_{\mu} \\ 
  \tilde{A}_\mu \\
  Z'_\mu \\
  \end{pmatrix}
  = R_{M}^T U_{2}^{-1} U_{1}^{-1} 
\begin{pmatrix}
Z_{\text{SM, } \mu} \\ A_\mu \\ K_\mu \\
  \end{pmatrix} \ .
  \label{eqn:BasisFinal}
\end{equation}
As a result of the basis change, $Z'_\mu$ now couples to SM fermions at tree level, while the coupling of the photon $\tilde{A}_\mu$ remains untouched due to the unbroken $U(1)_{EM}$ gauge symmetry.  

Moving beyond Ref.~\cite{Liu:2017lpo}, we are particularly interested in the double limit where the fractional mass splitting $\delta m \equiv (m_{Z,\text{ SM}} - m_K)/m_{Z,\text{ SM}}$ and kinetic mixing $\chi$ both approach $0$.  As evidenced by~\cref{eqn:allRots}, these limits do not commute, meaning the order in which the limits are taken is crucial.  If we first hold $\chi$ fixed and finite and take $\delta m \to 0$, then we arrive at a degenerate $Z$-$Z'$ system that has a maximal mixing with $\theta_M \to \pi/4$.  If we instead keep $\delta m$ fixed and finite and take $\chi \to 0$, then we obtain $\theta_M \to 0$ and the $Z$-$Z'$ system is again quasi-degenerate for small $\delta m$ but the $Z'$ boson is decoupled from SM currents.  The incompatibility of these limits demonstrates that the typical naive factorization of resonances fails in the degenerate regime, and more puzzling, approaching a $Z_2$ symmetric limit seems ill-defined.

The primary pitfall in this line of reasoning is that a consistent 1-loop treatment of the propagator matrix, as presented in~\cref{sec:Framework} will introduce 1-loop contributions to the mass matrix that will void the $\theta_M$ definition.  As a result, the inclusion of the pole masses extracted from the improved Breit-Wigner treatment also repackages the large mixing, shifting the interaction dominantly into one of the mass eigenstates.  Therefore, we justify the decoupling of the $Z'$ boson even when it is quasi-degenerate with the $Z$ boson.  Importantly, there is a regime in the kinetic mixing vs.~$Z'$ mass plane that is inaccessible because of avoided crossing in the tree-level $Z$-$Z'$ masses.  The derivation of the physics of the double limit is detailed in Appendix~\ref{sec:doublelimit}, where we also critically evaluate earlier literature that addressed the quasi-degenerate $Z$-$Z'$ regime.  

We are studying the competition of the explicit breaking of the $Z_2$ symmetry from 1-loop shifts in the pole masses versus the possible soft breaking from a tree-level mass splitting.  To understand the decoupling, we recognize that the mass squared matrix is symmetric but not Hermitian, and that the imaginary terms in the eigenvalues of the mass squared matrix will mark the degree of decoupling.  As similar observation in the context of scalar mixing with the Higgs boson was made in Ref.~\cite{Sakurai:2022cki}. 
To that end, we can formalize these subtleties about decoupling and mixing in the quasi-degenerate regime, where we dub the mechanism as the "dispersive seesaw" effect.

\subsection{A Toy Model to Illustrate the Dispersive Seesaw Effect}
\label{subsec:ToyModel}


To understand the dispersive seesaw mechanism, consider the $2 \times 2$ matrix
\begin{equation}
    \mathbf{M} = \mathbf{M}_{\mathrm{tree}} + \mathbf{M}_{\mathrm{1\,loop}} = 
    \begin{pmatrix}
        M^2 - i M \Gamma & - i \epsilon M \Gamma \\
       - i \epsilon M \Gamma & M^2 + \Delta_{\text {tree}} - i \epsilon^2 M \Gamma \\
    \end{pmatrix}
    \label{eqn:Mtoy}
\end{equation}
as a toy model of two particles of mass $M$ (\textit{i.e.}~$\mathbf{M}_{\mathrm{tree}} = \text{ diag} \left(M^2, \, M^2 + \Delta_{\text {tree}} \right) $) having identical interactions, but with the second particle's coupling constant different by a factor $\epsilon$ w.r.t~the first. We neglect the real part of the 1-loop self-energy because we are only interested in understanding the behavior of the imaginary part of the pole masses.  By the optical theorem, the imaginary piece of each entry of $\mathbf{M}$ is related to a tree-level scattering process, but because of mixing, the identification of an imaginary term in the self-energy with the width of any physical eigenstate is incorrect, as we will see below.  The absence of an imaginary piece is a self-consistency requirement for the corresponding vector to decouple.

First, if we assume $\Delta_{\text {tree}} \gg M\Gamma$, {\it i.e.}~large mass splitting, the eigenvalues are 
\begin{align}
    M_1^2 &= M^2  - i M \Gamma
    + \mathcal{O}\left( M \Gamma /\Delta_{\text {tree}}\right)  \ , \quad
    M_2^2 = M^2 + \Delta_{\text {tree}} - i\epsilon^2 M\Gamma
            + \mathcal{O}\left(M \Gamma/\Delta_{\text {tree}}\right) \ ,
\label{eqn:EigLargeSplit}
\end{align}
and we see that the second particle's width is equal to the first one but rescaled by $\epsilon^2$. In this regime, the usual Breit-Wigner approximation and factorization of resonances is completely justified.

However, if we assume perfect degeneracy, {\it i.e.}~$\Delta_{\text {tree}} =0$, the exact eigenvalues of this matrix are
\begin{align}
    M_1^2 &= M^2 - i (1 + \epsilon^2) M \Gamma \ , \quad
    M_2^2 = M^2 \ .
    \label{eqn:Delta0}
\end{align}
We see that one of the particles absorbs all the 1-loop effects, where the imaginary part of its two-point correlation function is the typical $- i M \Gamma$ plus an extra $\epsilon^2$ amount due to mixing.  The other particle, however, is left untouched by 1-loop renormalization, and in particular develops no width and remains stable, just like in the $\epsilon \to 0$ limit.
The key point here is that for $\Delta_{\text {tree}} = 0$, the tree-level mass matrix enjoys a $Z_2$ symmetry that is broken at 1-loop. But the breaking is saturated, in the sense that only one particle shows the breaking and the other remains at the previously $Z_2$ conserving value.

Finally, we analyze~\cref{eqn:Mtoy} for $\Delta_{\text {tree}} \ll M \Gamma$, where we have competition between the explicit soft breaking of the $Z_2$ symmetry by $\Delta_{\text {tree}}$ and the 1-loop $Z_2$ breaking from interactions.  The eigenvalues of $\mathbf{M}$ are now
\begin{align}
    M_1^2 &= M^2 - i (1 + \epsilon^2) M \Gamma
    + \dfrac{\epsilon^2}{1 + \epsilon^2} 
    \left(
    \Delta_{\text {tree}} +  \Delta_{\text {tree}}^2 \dfrac{-i  M \Gamma}
    {(1 + \epsilon^2 )^2 \left( M \Gamma\right)^2} 
    \right)
    + \mathcal{O}\left(\frac{\Delta_{\text {tree}}^3}{(M \Gamma)^2}\right) \ , \nonumber \\
    M_2^2 &= M^2 +
           \dfrac{1}{1 + \epsilon^2}
           \left( 
           \Delta_{\text {tree}} - \Delta_{\text {tree}}^2
           \dfrac{- i \epsilon^2  M \Gamma}
           {(1 + \epsilon^2 )^2 \left( M \Gamma\right)^2} 
           \right)
            + \mathcal{O}\left( \frac{\Delta_{\text {tree}}^3}{(M \Gamma)^2} \right) \, .
\label{eqn:QZEigZ2break}
\end{align}
Importantly, the imaginary part of $M_2^2$ is now suppressed by $(\Delta_{\text {tree}}/M\Gamma)^2$ with respect to the large mass splitting case in~\cref{eqn:EigLargeSplit}, and moreover, the first particle still acquires the dominant $(1+ \epsilon^2) M \Gamma$ imaginary term as in~\cref{eqn:Delta0}.  Indeed, the width of the second particle is
\begin{equation}
 \Gamma_2 = -\frac{\Im(M_2^2)}{M} \approx
 \epsilon^2 \, \Gamma  \left( \dfrac{\Delta_{\text {tree}}}{M \Gamma}\right)^2 \ll \epsilon^2 \, \Gamma\, ,
 \label{eqn:Gamma2}
\end{equation}
where we further assumed $\epsilon \ll 1$, demonstrating explicit agreement with Ref.~\cite{Sakurai:2022cki}.

We now apply the features of the dispersive seesaw effect and the aforementioned avoided crossing behavior from~\cref{eqn:MassEig} to our quasi-degenerate $Z'$ study.

\subsection{Applying the dispersive seesaw effect to the quasi-degenerate $Z'$ regime}
\label{subsec:QZapplied}

In our $Z$-$Z'$-$\gamma$ scenario with kinetic mixing $\chi$ and fractional mass splitting $\delta m$ between gauge basis vectors $Z_{\text{SM, } \mu}$ and $K_\mu$, the width of the $\tilde{Z}_\mu$ mass eigenstate changes from $\cos^2{\theta_M}\Gamma_{Z,\text{ SM}}$ to $\Gamma_{Z,\text{ SM}} \left(1 + \left(\tan{\theta_M} m_{\tilde{Z}},\delta m^{\text{Phys}} / \Gamma_{Z,\text{ SM}} \right)^2\right)$
after diagonalizing the 1-loop mixing, where we defined the fractional mass splitting of the physical eigenstates $\tilde{Z}$ and $Z'$,
\begin{equation}
\delta m^{\text{Phys}} = (m_{\tilde{Z}} - m_{Z'} ) / m_{\tilde{Z}}
\end{equation}
and $\theta_M$ is the final rotation angle of~\cref{eqn:allRots}. On the other hand, the width of the $\tilde{K}_\mu$ mass eigenstate changes from $\sin^2{\theta_M} \, \Gamma_{Z,\text{ SM}}$ to $\tan^2{\theta_M}\, \Gamma_{Z,\text{ SM}} \left( m_{\tilde{Z}} \, \delta m^{\text{Phys}} / \Gamma_{Z,\text{ SM}} \right)^2$, as a consequence of the dispersive seesaw effect.  Given $m_{\tilde{Z}} \, \delta m^{\text{Phys}} \ll \Gamma_{Z,\text{ SM}}$, the $\tilde{K}_\mu$ vector exhibits decoupling, since its contribution to $Z$-like scattering processes is suppressed by $\tan^2{\theta_M} (m_{\tilde{Z}} \, \delta m^{\text{Phys}} / \Gamma_{Z,\text{ SM}})^2$.  Again, this arises because the diagonalization of the 1-loop propagator matrix has removed the $\chi$-induced mixing between the vector states and shifted the leading $\tan^2{\theta_M} \, \Gamma_{Z,\text{ SM}}$ effect into the $\tilde{Z}$ propagator.

We remark that the reverse logic self-consistently describes the effect of a possible $\tilde{Z}$ contribution to $K$-like amplitudes, where we can introduce a dark current and $\Gamma_K$ width for the $K_\mu$ vector.  Here, the kinetic mixing leads to a dispersive seesaw effect on the $\Gamma_K$ width correction to the imaginary piece of the $\tilde{Z}$ two-point correlation function, which will exhibit the same $\tan^2{\theta_M} (m_{Z'} \, \delta m^{\text{Phys}} / \Gamma_{K})^2$ suppression.

\begin{figure}[htb!]
\includegraphics[width=0.49\textwidth]{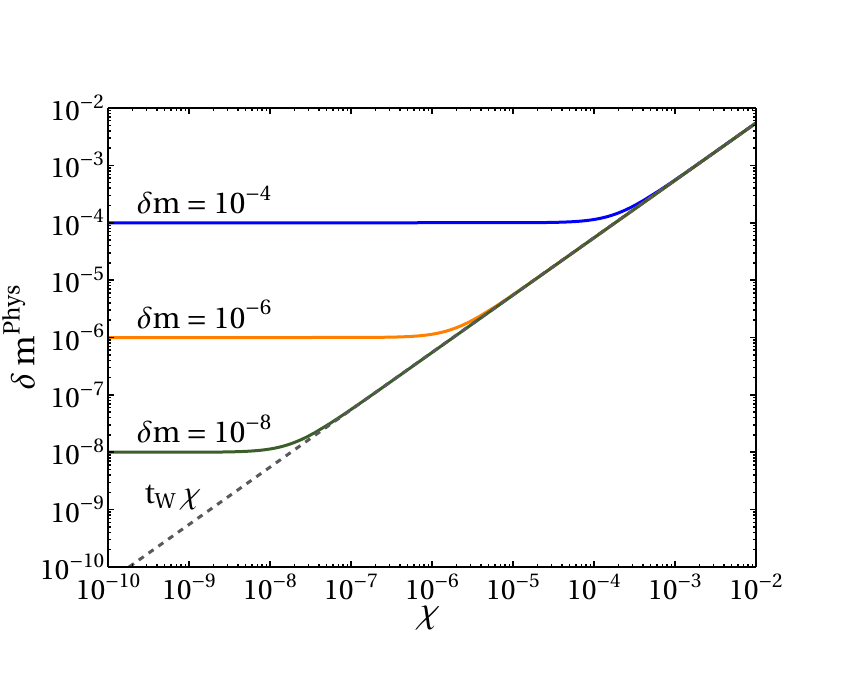}
\includegraphics[width=0.49\textwidth]{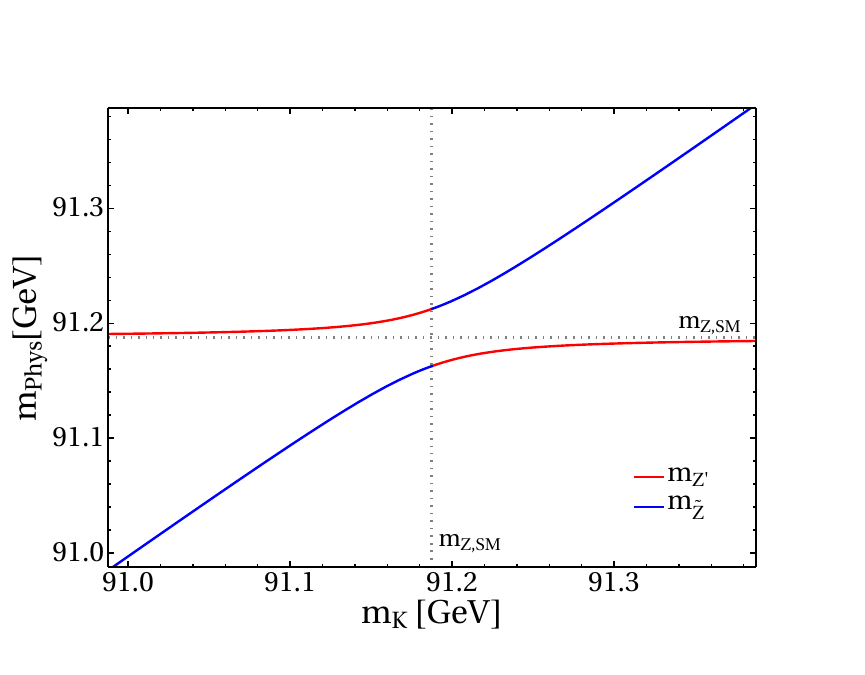}
\caption{Left panel: Physical mass splitting as a function of the kinetic mixing $\chi$ for three different values of the input fractional mass splitting $\delta m$.  When $\chi \ll |\delta m|$, the physical mass splitting is equal to the input one, while for $\chi \gg |\delta m|$, the physical mass splitting follows a universal linear scaling with slope $t_W$.  Right panel: Avoided crossing of the mass eigenvalues for fixed $\chi = 10^{-3}$ as a function of input mass $m_K$.}
\label{fig:delPhysAvoided}
\end{figure}

We stress that, because of avoided crossing, the tree-level eigenvalues from~\cref{eqn:MassEig} are never the same for any non-vanishing $\chi$, and hence including the 1-loop correction will always induce a small but suppressed width for the $Z'$ boson.  The avoided crossing also dictates that the kinetic mixing $\chi$ and the physical mass splitting between the $\tilde{Z}$ and $Z'$ bosons are not independent, as shown in~\cref{fig:delPhysAvoided}.  Namely, for a fixed $\chi$, $\delta m^{\text{Phys}}$ has two asymptotic limits:
\begin{equation}
\begin{cases}
\delta m^{\text{Phys}} = t_W \chi \quad \text{for} \quad 
|\delta m|  \ll \chi \ , \\
\delta m^{\text{Phys}} = \delta m \quad \text{for} \quad \chi \ll |\delta m| \ll 1 \ . 
\label{eqn:deltamPhys}
\end{cases}   
\end{equation}
Hence, for fixed $\chi$, the physical mass $m_{Z'}$ cannot be arbitrarily close to $m_{Z,\text{ SM}}$, and instead we always retain $t_W \chi$ as a dimensionless order parameter of the mass splitting,  This is illustrated in the right panel of~\cref{fig:delPhysAvoided} for $\chi = 10^{-3}$, where varying the input $m_K$ parameter smoothly through $m_{Z, \text{ SM}}$ retains a nonzero $\delta m^{\text{Phys}}$ for the entire range.

The following approximation is valid in the $\chi \ll \abs{\delta m} \ll 1$ limit:
\begin{equation}
    m_{Z'} \approx m_{Z,\, \text{SM}} (1 \pm t_W \chi / 2) \ ,
\end{equation}
the upper (lower) sign corresponding to $\delta m > 0$ ($\delta m < 0$).

Moreover, as applied to the $Z$ pole, a given $\chi$ dictates a minimum $\Delta_{\text{tree}}$, and therefore, the width suppression from the dispersive seesaw effect~\cref{eqn:Gamma2} cannot completely decouple the $Z'$ boson.

\section{Collider Constraints on Kinetic Mixing}
\label{sec:results}

In this section, we apply the formalism presented in~\cref{sec:Framework} to the $Z'$ model discussed in~\cref{sec:Model}, to derive collider constraints on the kinetic mixing parameter $\chi$ in the mass region where the $Z'$ boson is quasi-degenerate with the $Z$ boson.  

We eschew performing a full fit to electroweak precision data and instead focus on three observables that would dominate such a fit: the mass of the $W$ boson, the width of the $Z$ boson, and the lineshape for muon pair production ($e^+ e^- \to \mu^+ \mu^-$) measured by the LEP experiments~\cite{ALEPH:2005ab}.  These observables are calculated using the six model parameters $m_{Z,\text{ SM}}$, $m_K$, $s_W$, $\chi$, $\Gamma_{Z,\text{ SM}}$ and a possible $\Gamma_K$.  As noted in~\cite{Babu:1997st}, $s_W$ in the on-shell scheme admits a flat direction in the $(m_{Z,\text{ SM}}, s_W)$ direction of the global electroweak fit, so we set $s_W$ self-consistently with the $m_{\tilde{Z}}$ output of our calculations.

We use the $W$ boson mass instead of the $Z$ boson mass as our observable because the standard global electroweak fit treats the $Z$ mass as an input, testing consistency to the SM electroweak parameters by allowing the input $Z$ mass parameter to float within its $\sim 2$ MeV uncertainty.  For our purposes, it is more tractable to use the aforementioned model parameters and test the consistency with the measured $W$ mass, which is defined in the global SM electroweak fit as~\cite{ParticleDataGroup:2024cfk}
\begin{align}
     m_W &= m_Z \, c_W(m_Z) \, \rho^{1/2} \ , 
     \label{eqn:mWmZrelation} \\
     \rho^{1/2} &= 1.01019 \pm 0.00009 \ ,
\end{align}
where the running of the weak angle has been absorbed into the measured values of the Fermi constant $G_F$ and cosine of the weak mixing angle at $m_Z$, $c_W (m_Z)$, while the $\rho$ parameter absorbs the remaining dependence on the relatively poorly-measured top and Higgs masses.  

\begin{figure}
    \centering
    \includegraphics{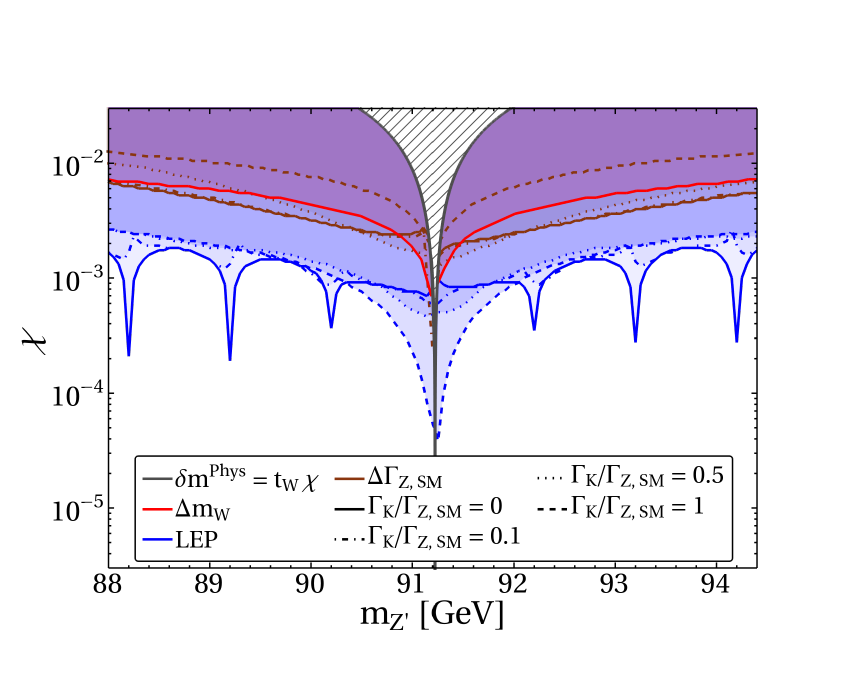}
    \caption{Constraints in the $(m_{Z'}, \, \chi)$ plane arising from excluded shifts in the $W
    $ mass (red line) and $Z$-like width (brown lines), as well as muon pair production (blue lines) from LEP Run 1~\cite{ALEPH:2005ab}.  The gray hatched region labeled ``$\delta m^{\text{Phys}} = t_W \, \chi$" shows the $(m_{Z'}, \, \chi)$ forbidden region from avoided crossing.}
    \label{fig:chivsmZp}
\end{figure}

By employing this definition, we assume there is no tension in the observed values of the electroweak precision observables and the observed $m_W$\footnote{We remark that the PDG average of $m_W$ determinations~\cite{ParticleDataGroup:2024cfk}, excluding the CDF Run-II result~\cite{CDF:2022hxs}, does indicate a mild 1.6$\sigma$ tension.  We thank an anonymous referee for informing us.}, so we will employ the world average of $m_W = 80.377 \pm 0.012$ GeV~\cite{ParticleDataGroup:2024cfk}, which does not include the recent CDF Run-II result~\cite{CDF:2022hxs}. The better precision of the new CDF result would not appreciably change the magnitude of the constraints on $\chi$.  Following our framework, we calculate the $Z$-like mass eigenvalue from the real part of the pole of the Breit-Wigner propagator, and use~\cref{eqn:mWmZrelation} to test consistency with the experimental $m_W$ value.  This is shown as the $\Delta m_W$ line in~\cref{fig:chivsmZp}.

The second major effect from the $Z'$ model would be a modification of the observed total $Z$ width, $\Gamma_{\tilde{Z}}$, from the imaginary part of the pole of the Breit-Wigner propagator.  The constraints coming from $\Gamma_{\tilde{Z}}$ moderately depend on the $\Gamma_K$ parameter that would characterize the effect of a dark vector current coupling for the $K$ boson.  As noted in~\cref{sec:Model}, the modification of the observed $Z$ width by finite $\Gamma_K$ is suppressed by the dispersive seesaw effect, a feature which has been missed in previous literature, as described in~\cref{sec:doublelimit}.  The constraints from the modification of the $Z$ width are shown as $\Delta \Gamma_{Z, \,\text{ SM}}$ lines in~\cref{fig:chivsmZp} for given choices of $\Gamma_K / \Gamma_{Z,\text{ SM}} = 0$, $0.1$, $0.5$, and $1$.  We advocate that a modern global fit focusing on kinetic mixing constraints in the quasi-degenerate regime should use a propagator matrix structure as described in~\cref{sec:Framework}, since the one-particle Breit-Wigner propagator form would miss essential features of the physics of the double resonance.

\begin{figure}[htb]
    \centering
    \includegraphics[width=0.7\textwidth]{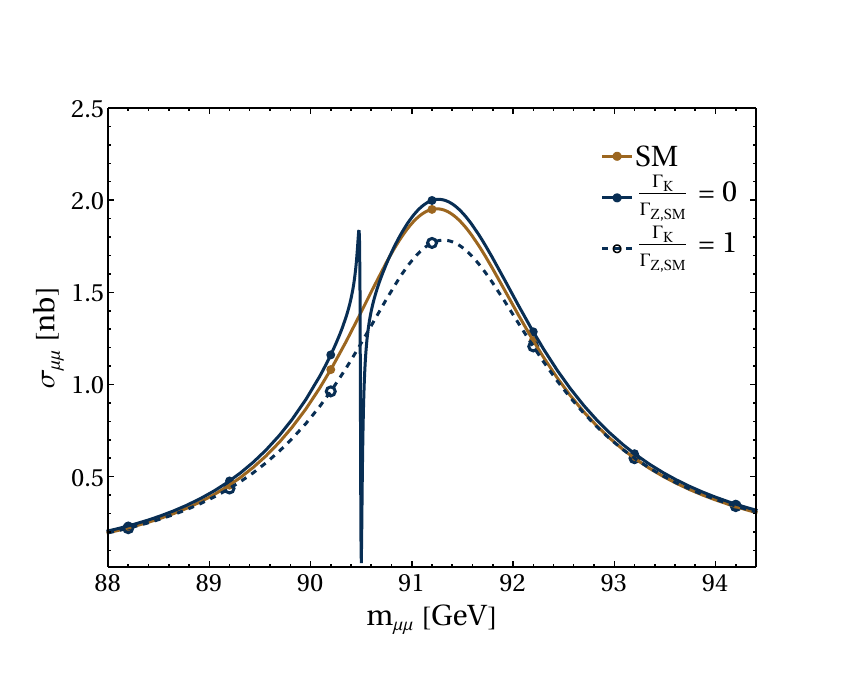}
    \caption{Cross section rates showing the $Z'$ resonance peak on top of the $Z$ peak in muon pair production for $m_{Z'} = 90.5$~GeV, $\chi = 5 \times 10^{-3}$, and either $\Gamma_K = 0$ (solid blue, filled blue circles) or $\Gamma_K = \Gamma_{Z, \text{SM}}$ (dashed blue, hollow blue circles).  The SM prediction is also shown (brown line, filled brown circles).  The circles indicate the LEP $\sqrt{s}$ energies used in Run 1.}
    \label{fig:SharpPeak}
\end{figure}

This last point also justifies the importance of considering the discretized lineshape of muon pair production as measured by the LEP experiments~\cite{ALEPH:2005ab}.  We adopt the specific $\sqrt{s}$ center of mass (c.o.m.) energies used in Run 1 of the LEP collider to deconvolute the extracted $Z$ resonance lineshape from a possible sharp resonance feature from a $Z'$ boson.  This is illustrated in~\cref{fig:SharpPeak}, which shows the comparison between the continuous versus discrete $m_{\mu \mu}$ distribution of a $Z'$ boson with $m_{Z'} = 90.5$~GeV and $\chi = 5 \times 10^{-3}$ and $\Gamma_K = 0$ or $\Gamma_K = \Gamma_{Z, \text{ SM}}$.  Clearly, the sharp feature is missed by the discrete scan in $\sqrt{s}$, especially given the beam energy resolution of 0.15\% at LEP Run 1~\cite{Martinez:1998rs}, and the resulting discretized points only show a relatively small deviation from the SM expectation.  Importantly, the lineshape of muon pair production deviates more strongly for $\Gamma_K = \Gamma_{Z, \text{ SM}}$, since the $Z'$ feature is distributed over a larger energy range and hence illustrates the importance of calculating the lineshape dependence on $\Gamma_K$ accurately.

To calculate the constraints on $m_{Z'}$ and $\chi$, we compute the muon pair production cross section on a grid of parameter points for the kinetic mixing model as well as the $\chi \to 0$ and $m_K \to \infty$ decoupled limit that corresponds to the SM.  We sample the cross section rate with the LEP $\sqrt{s}$ scan points for a total of $10^6$ muon pairs using a Gaussian beam energy spread $0.15\%$~\cite{Martinez:1998rs}.  We then consider the absolute sum of the relative residual defined by these rates, 
\begin{equation}
\Delta_{\mu\mu}  = \sum_{\sqrt{s}\in \, \text{LEP scan}} \abs{\frac{\sigma^{\text{SM}+Z'}_{\mu\mu}(\sqrt{s})-\sigma^{\text{SM}}_{\mu\mu}(\sqrt{s})}{\sigma^{\text{SM}}_{\mu\mu}(\sqrt{s})}}\, ,
\end{equation}
and compare $\Delta_{\mu\mu}$ to the combined statistical and systematic uncertainty from the LEP experiments: this combined uncertainty is estimated from the peak rate of muon pair production,
\begin{equation}
\sigma^{\text{SM}}_{\mu\mu}(\sqrt{s} = m_{Z,\text{ SM}}) = \frac{12\pi}{m_{Z,\text{ SM}}^2}\frac{\Gamma^{ee}_{Z,\text{ SM}}\Gamma^{\mu\mu}_{Z,\text{ SM}}}{\Gamma_{Z,\text{ SM}}^2}\, ,
\label{eqn:SigmaPeak}
\end{equation}
where $\Gamma^{ee}_{Z,\text{ SM}}(\Gamma^{\mu\mu}_{Z,\text{ SM}})$ is the partial width of the SM $Z$ to $e^+e^-(\mu^+\mu^-)$. Hence the experimental uncertainties on the values of the global SM electroweak fit combine to give the aggregate uncertainty for the muon pair production cross section.  After including the correlations between the quantities in~\cref{eqn:SigmaPeak}, we impose, at 90\% C.L.,
\begin{equation}
   \Delta_{\mu\mu} < 0.76\% \ .
\end{equation}
The resulting constraints are shown as the blue lines in~\cref{fig:chivsmZp} for various choices of $\Gamma_K$.  As seen in~\cref{fig:chivsmZp}, the larger $\Gamma_K$ choices typically lead to stronger constraints, with the notable exception of the spikes for $\Gamma_K = 0$.  Here, the fortunate coincidence between the $Z'$ mass and the $\sqrt{s}$ energy from LEP Run 1 lead to enhanced constraints.

In~\cref{fig:chivsmZp}, we also show a gray hatched region to indicate the forbidden quasi-degenerate regime arising from avoided crossing, which was illustrated in~\cref{fig:delPhysAvoided}.  For a given $\chi$, the avoided crossing forces the physical $m_{Z'}$ to deviate from $m_{Z, \text{ SM}}$ by a minimum amount, see~\cref{eqn:deltamPhys}.

\begin{figure}
    \centering
    \includegraphics[width=0.7\textwidth]{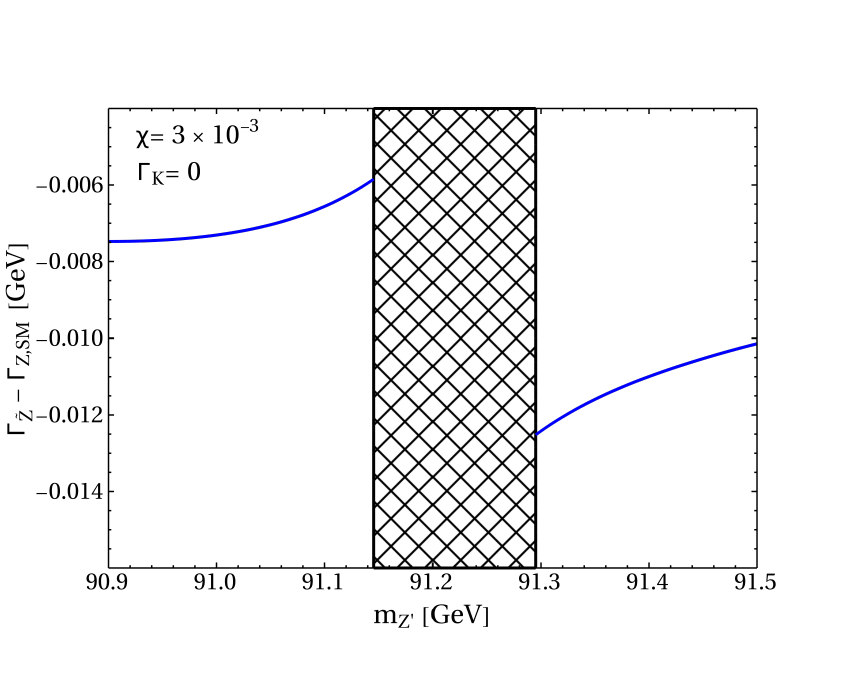}
    \caption{Shift in the SM-like $\tilde{Z}$ width, for an illustrative parameter point of $\chi = 3 \times 10^{-3}$, $\Gamma_K = 0$, as a function of the physical mass $m_{Z'}$. 
    The discontinuous jump of the width around $m_{Z'} = m_{Z,\text{ SM}}$ is evident.}
\label{fig:AvCrossGamma}
\end{figure}

We remark that the asymmetry in the constraints around the $m_{Z,\text{ SM}}$ line is a physical effect of the fact that the imaginary part of the pole masses are discontinuous around $m_{Z'} = m_{Z,\text{ SM}}$.  
This is shown in~\cref{fig:AvCrossGamma} for $\chi = 3 \times 10^{-3}$ and $\Gamma_K = 0$.  
This discontinuity is caused by the appearance of a mass threshold at $m_{\tilde Z}$: as $m_{Z'}$ is dialed to be larger than $m_{\tilde{Z}}$, the $Z' \to \tilde{Z}$ transition gets enhanced resulting in a larger width for the $Z'$ bosons, and a smaller width for the $\tilde{Z}$ boson, because the $\tilde{Z} \to Z'$ process is now classically forbidden.  Farther from the degenerate point but still in the $m_{Z'}-m_{Z,\text{ SM}} \lesssim \Gamma_{Z,\text{ SM}}$ region, this mass threshold is almost negligible and the constraints are approximately symmetrical around $m_{Z,\text{ SM}}$. We note here that this effect, which can only be captured if the energy dependence of the self-energies, is not displayed by the effective Hamiltonian treatment of~\cref{subsec:ToyModel}.

\begin{figure}
    \centering
    \includegraphics{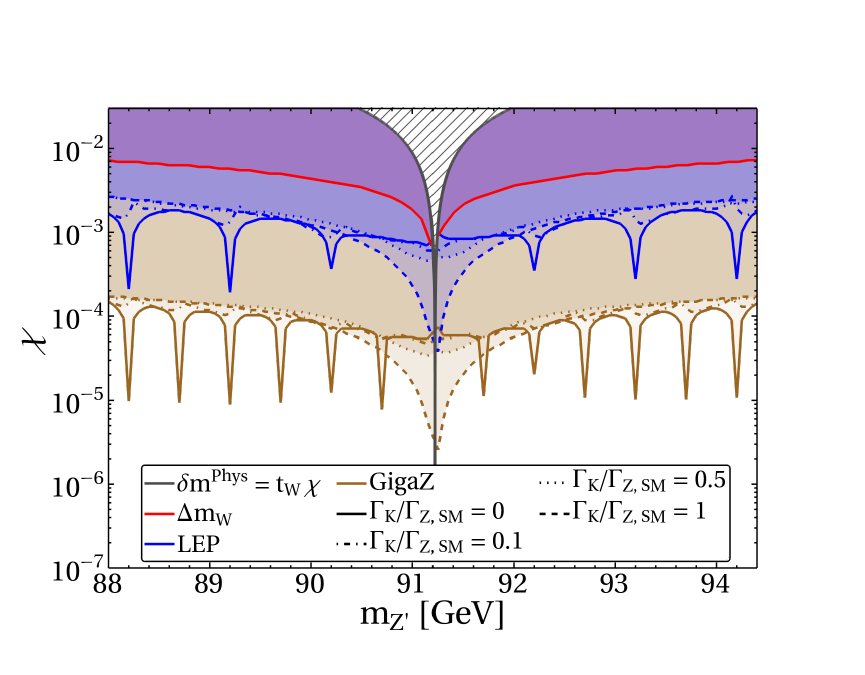}
   \caption{The same constraints in the $(m_{Z'},\chi)$ plane from Fig.~\ref{fig:chivsmZp}, with the projection for a Giga-$Z$ factory (light brown lines) running over twice as many $\sqrt{s}$ scan points compared to LEP Run 1.}
    \label{fig:GigaZPlot}
\end{figure}

Having emphasized the importance of the $\sqrt{s}$ collider parameter choices and the beam energy resolution in testing for a possible second resonance near the $Z$ pole, we estimate the projected sensitivity improvement in the $\chi$ vs.~$m_{Z'}$ plane from a GigaZ factory in~\cref{fig:GigaZPlot}.  For the GigaZ machine, we assume a finer energy scan, with twice the number of $\sqrt{s}$ choices as LEP Run 1, as well as an effective sensitivity reflecting the higher luminosity and assumed systematic uncertainties improving over the LEP experiments by two orders of magnitude.  We reserve a dedicated sensitivity study of the different future $e^+ e^-$ machines operating at the $Z$ pole, such as FCC-ee~\cite{FCC:2018evy} and CEPC~\cite{CEPCStudyGroup:2018ghi}, for future work.  

We also reserve a complete evaluation of the electroweak global fit adding the effects of the quasi-degenerate $Z'$ boson for future work, which would include observables such as forward-backward asymmetries and modifications of $\alpha_{\text{EM}}$.  One important aspect of such a global fit would be fixing the normalization of the observed $Z$ peak in comparison to the off-shell region, to avoid theoretical biases about additional resonances or other lineshape distortions.  This global fit would also give a robust result for the increase in the quoted $Z$ mass uncertainty of $\sim 2$~MeV when a quasi-degenerate $Z'$ boson is added to the theory.

\section{Conclusions}
\label{sec:conclusions}
We have derived consistent collider constraints on a kinetically mixed $Z'$ boson that is nearly degenerate with the SM $Z$ boson, emphasizing the use of an improved Breit-Wigner prescription to avoid an uncontrolled perturbative expansion. We analyze both model-independent constraints, characterized by the kinetic mixing parameter $\chi$, and model-dependent constraints, such as $Z'$ width effects arising from a possible coupling of the $Z'$ boson to a dark sector.  Our main results are shown in~\cref{fig:chivsmZp}, where we show the constraints from the LEP measurements in the $\chi$ vs.~$m_{Z'}$ plane for the quasi-degenerate regime.

To explain the features seen in~\cref{fig:chivsmZp}, we needed to consider a double limit of vanishing mass splitting and kinetic mixing and how this limit is calculated using the 1-loop improved Breit-Wigner framework from~\cref{sec:Framework}.  This motivated a natural explanation of the phenomenology in terms of the dispersive seesaw effect and avoided level crossing, which govern the imaginary and real parts of the poles of the Breit-Wigner propagators, respectively.  We also remarked in~\cref{sec:doublelimit} how these phenomena were missed in previous work, leading to wrong conclusions in the nearly degenerate region. Although our treatment via the 1-loop improved Breit-Wigner prescription allows us to derive the correct widths and to derive the dispersive seesaw effect, a prescription where the leading order width already approximates the 1-loop improved widths would be useful. In the current framework, the leading order widths undergo an order one correction by the 1-loop improvement.  This issue has been partially addressed in Ref.~\cite{Boyanovsky:2017esz}, in the context of an effective action for scalar mixing, as well as the very recent Ref.~\cite{Kamada:2024ntk}, in the context of vector boson mixing.  In contrast to Ref.~\cite{Kamada:2024ntk}, we explicitly calculate and show the collider constraints in the $\chi$ vs.~$m_{Z'}$ plane.  We highlight that our results show that the real and imaginary parts of the pole masses are discontinuous at $m_K \simeq m_{Z,\text{ SM}}$, in contrast to the claim in Ref.~\cite{Kamada:2024ntk} that the imaginary part remains continuous. 

While the phenomenological consequence of the dispersive seesaw effect and avoided crossing play an important role in extracting consistent constraints on $Z'$ bosons at 1-loop in the quasi-degenerate regime, we are also interested in the analysis of these phenomena at higher perturbative orders, which we reserve for future work.

We remark that although our motivation was extracting $Z'$ constraints in a degenerate mass window around the $Z$ boson from collider experiments, our framework serves as the basis for understanding novel phenomenology where two degenerate vectors are responsible for generating the dark matter, for example.  This is also related to studies of resonant leptogenesis~\cite{Pilaftsis:2003gt}.  Studying the novel cosmological phenomenology of two highly-mixed and quasi-degenerate vector bosons is also reserved for future work.  Also reserved for future is the possible connection between the dispersive seesaw effect demonstrated in the toy model studied in~\cref{subsec:ToyModel} and the quantum Zeno effect~\cite{Misra:1976by}, which is a feature of quantum systems that allows the time evolution of a given system to be arbitrarily slowed down by repeated measurements.

\section*{Acknowledgments}
\label{sec:acknowledgments}

This research is supported by the Cluster of Excellence PRISMA$^+$, ``Precision Physics, Fundamental Interactions and Structure of Matter" (EXC 2118/1) within the German Excellence Strategy (project ID 390831469).  The authors would like to thank the Fermilab theory group for their hospitality in the completion of this work.  The authors would like to thank Susan Gardner for illuminating discussions and Anke Biek\"{o}tter for helpful discussions about $Z$ pole physics.

\appendix
\section{The double limit of vanishing mass difference and kinetic mixing}
\label{sec:doublelimit}

In this Appendix, we discuss the physics of the double limit for $Z'$ bosons that are kinetically mixed with hypercharge and also quasi-degenerate with the $Z$ boson.  The corresponding $\chi \to 0$ and $\delta m = (m_{Z,\text{ SM}} - m_K)/m_{Z,\text{ SM}} \to 0$ limits do not commute, which we demonstrate using the definition of $\beta$ and $\tan \theta_M$ in~\cref{eqn:allRots}.

If we first keep $\chi$ finite and send $\delta m \to 0$, then $\beta \propto \chi$ and sending $\chi \to 0$ will give $\theta_M \to \pi/4$, leading to a degenerate $Z-Z'$ system that is maximally mixed.  On the other hand, if we keep $\delta m$ finite and take $\chi \to 0$, then $\beta \to \infty$ and $\theta_M \to 0$, which causes the $Z'$ boson to decouple completely from the SM.

We remark that, unfortunately, previous work~\cite{Hook:2010tw} did not correctly analyze the physics of this double limit.  Concretely, Ref.~\cite{Hook:2010tw} uses the definition of the $Z-Z'$ mixing angle from Ref.~\cite{Cassel:2009pu}, which we express in our notation as
\begin{equation}
\tan{\theta_M} = 
    \dfrac
    {2m_{Z,\text{ SM}}^2 t_W \dfrac{\chi}{\sqrt{1-\chi^2c_W^{-2}}}\left(m_{Z'}^{2}- m_{Z,\text{ SM}}^2 \right)}
    {\left(m_{Z'}^{2}-m_{Z,\text{ SM}}^2 \right)^2 - \left(m_{Z,\text{ SM}}^2 t_W \dfrac{\chi}{\sqrt{1-\chi^2c_W^{-2}}}\right)^2 } \ .
    \label{eqn:WrongMassShift}
\end{equation}
Unlike~\cref{eqn:allRots}, this definition of $\tan \theta_M$ uses the physical $Z'$ mass instead of the input $m_K$ parameter.  If this distinction is neglected and we incorrectly replace $m_{Z'}^{2}$ above by $m_K^2$, then the mixing angle appears to cross zero at $m_K = m_{Z, \, \text{SM}}$, leading to the conclusion that the exactly mass-degenerate limit generates no shift in the mass.  The appearance of a zero in the mass shift expression by~\cite{Hook:2010tw} is contrary to the phenomenon of avoided crossing.

We remark that the $\Gamma_{\tilde{Z}}$ constraints in~\cite{Hook:2010tw} are also inconsistent as a result of using the large mass-splitting approximation for the invisible width of the $\tilde{Z}$ boson,
$    \Gamma_{\tilde{Z},\,\text{inv}} \approx \tan^2{\theta_M}\, \frac{m_{Z,\text{ SM}}}{m_{Z'}} \Gamma_K $.
As we explain in~\cref{sec:Model}, while this rescaling of widths is appropriate for large mass-splitting, the qualitatively different behavior of the dispersive seesaw effect takes over for the quasi-degenerate regime, where the $\Gamma_K$ width gives a highly suppressed contribution to the invisible width of the $\tilde{Z}$ boson. Our results from the change in the $Z$ width are calculated using the 1-loop improved BW prescription, and are shown~\cref{fig:chivsmZp}.


Another feature of the treatment in~\cite{Hook:2010tw} is a saw-tooth shape of the kinetic mixing constraints in the quasi-degenerate region, where collider experiments are used to give constraints on $Z'$ masses continuously across the $Z$ pole.  Although our consistent constraints also exhibit a sawtooth shape, this shape arises because of avoided crossing.  Hence, for a given $\chi$, we cannot populate $Z'$ masses  continuously across the $Z$ pole, which is another aspect of the physics of the quasi-degenerate regime absent in Ref.~\cite{Hook:2010tw}.




\bibliographystyle{JHEP}
\bibliography{Refs}

\end{document}